\documentclass[11pt,reqno]{amsart}

\usepackage{enumerate}
\usepackage{graphicx,graphics}
\usepackage{soul}
\usepackage{amsfonts}
\usepackage{amssymb}
\usepackage{amsthm}
\usepackage{amsmath}
\usepackage{mathrsfs,color}
\usepackage{enumitem}
\usepackage{subcaption}
\usepackage{varwidth}
\usepackage{multirow}
\usepackage{hhline}
\usepackage{hyperref}

\newcommand{\abs}[1]{\left\lvert#1\right\rvert}
\newcommand{\R}{\mathbb{R}}

\newtheorem{theorem}{Theorem}[section]
\newtheorem{proposition}[theorem]{proposition}
\newtheorem{corollary}[theorem]{corollary}

\theoremstyle{definition}
\newtheorem{definition}[theorem]{Definition}
\newtheorem{example}[theorem]{example}
\newtheorem{remark}[theorem]{remark}

\title{Moduli of Continuity in Metric Models and Extension of Liveability Indices.}

\author{Roger Arnau, J. M. Calabuig, Álvaro González and Enrique A. Sánchez Pérez}

\address{ 
	Instituto Universitario de Matem\'atica Pura y Aplicada,
	Universitat Polit\`ecnica de Val\`encia, Camino de Vera s/n, 46022
	Valencia, Spain.}
\email{
	ararnnot@posgrado.upv.es,
	jmcalabu@mat.upv.es,
	agoncor@posgrado.upv.es,
	easancpe@mat.upv.es}

\begin{document}

\maketitle
	
\begin{abstract}
	Index spaces serve as valuable metric models for studying properties relevant to various applications, such as social science or economics. These properties are represented by real Lipschitz functions that describe the degree of association with each element within the underlying metric space. After determining the index value within a given sample subset, the classic McShane and Whitney formulas allow a Lipschitz regression procedure to be performed to extend the index values over the entire metric space. To improve the adaptability of the metric model to specific scenarios, this paper introduces the concept of a composition metric, which involves composing a metric with an increasing, positive and subadditive function $\phi$. The results presented here extend well-established results for Lipschitz indices on metric spaces to composition metrics. In addition, we establish the corresponding approximation properties that facilitate the use of this functional structure. To illustrate the power and simplicity of this mathematical framework, we provide a concrete application involving the modelling of livability indices in North American cities.
\end{abstract}

\section{Introduction}
Metric models, constructed from the aggregation of various variables, serve as useful frameworks that enable prospective approaches in fields such as the social sciences. One version of these models is given by so-called index spaces (see \cite{index} and related literature) which consist of a triple $(M,d,I)$, where $(M,d)$ represents a metric space and $I$ denotes an index (essentially a non-negative Lipschitz function satisfying additional regularity properties). Such models prove advantageous in practical contexts, as they allow the extension of significant indices defined within a metric subspace of $(M,d)$ to the entire space, while preserving the Lipschitz constant, as seen in the classical scenario of Lipschitz regression. This methodology is justified when attempting to model special circumstances that do not conform to linear constraints. The usefulness of these results has been demonstrated in recent years by a large number of research papers in various disciplines. For example, concrete applications can be observed in machine learning, where this conceptual framework is employed \cite{machine}. Analogous concepts are also widely used in other scientific fields \cite{memoli2006, dacorogna2006}.

However, there are many cases where the metric of the problem is not established and has to be defined at some stage in the modelling process. This choice can be critical, as different metrics can lead to very different results. This article attempts to redefine this methodology by introducing a novel approach to obtain, in a straightforward manner, from a simple metric, another metric that better fits the problem. This newly proposed metric, denoted $d_{\phi}$, is chosen from the set of composition metrics $\phi \circ d$, where $d$ is the original metric and $\phi$ is a continuity modulus. In particular, we show that this adaptation of the initial metric space yields significant improvements over using the original metric structure $(M,d)$. To this end, we provide an example consisting in extending the AARP livability index, known for certain US cities, to a wider range of cities, thereby evaluating and analysing the errors incurred under different considerations. Furthermore, we introduce a new category of indices, called \textit{standard indices}, aimed at increasing the effectiveness of the proposed approximation technique. In all these cases, the Lipschitz constant, together with other normalisation properties related to the Katetov condition, serves as a central tool to control the resulting extension over all scenarios.

The ideas presented here are not new. The concept of composition 
metrics can be found, under various names, in several works, 
including some articles from the early part of the twentieth century. 
The notion of continuous metric transformations (metric preserving 
functions in the literature) is first introduced in Wilson's 1935 
article \cite{wilson1935}. A paper by Bors\'ik and Dobo\v{s} in 1981 \cite{dobos1981} compiles some of the conditions under which a given function composed with a metric will return a metric, and vice versa. Furthermore, the article by Valentine \cite{valentine1943} already introduces a continuity modulus in the Lipschitz condition.

Therefore, our interest is to carry out all this study in the framework of metric models related to the extension of Lipschitz functions. Specifically, by introducing a new class of Lipschitz functions that aims to generalise this notion by incorporating an alternative metric for improved extension capabilities. This involves a more skilful adaptation to the data set and an improved extension of the indices. Our research will explore the properties and instances of this function space that are relevant to our focus. We will explore its theoretical properties and relevance to the field of interest, in particular index theory, while analysing its effectiveness and contrasting it with established methods. To achieve this, we have organised this paper as follows. In Section \ref{sec:basic} we will present the main concepts on real Lipschitz functions and metric spaces, and also the generalisation proposal we have introduced (composition metrics and other mathematical tools). The Section \ref{sec:indexspaces} will focus on the mathematical framework on which the indices and the corresponding rankings are based. We will study index spaces based on the new class of functions introduced, and we will present two techniques that allow us to extrapolate the information provided by the indices for elements in which it is not defined. Thus, we will provide possible approximations for these elements, which will be the main point of interest. Finally, in section \ref{sec:apliccation} we will explore the practical application of the above findings to algorithms that numerically execute the described extension procedures. We will delve into specific examples where we examine their functionality, paying particular attention to performance differences compared to the original alternative and other potential procedures.

\section{Basic definitions and concepts}\label{sec:basic}

We begin this section by recalling results and basic concepts related to Lipschitz maps, having previously established the theoretical framework in which they are found: metric spaces. The work of Deza et al. \cite{distancias} and Cobzaş et al. \cite{lipschitzbook} has been followed for this part.

    If $D$ is a nonempty set and $d\colon D\times D\to\R^+$, where $\R^+$ is the set of non-negative real numbers, $d$ is a \textit{distance} or \textit{metric} if it satisfies:
$d(a,b)=0 \ \text{if and only if} \ a=b$,
        $d(a,b)=d(b,a)$ and
        $d(a,b)\leq d(a,c)+d(c,b) \ \text{for all} \ a,b,c \in D$.
The pair $(D,d)$ is called \textit{metric space}, and for any subset $D_0\subseteq D$ with the distance restricted to it, $d\restriction_{D_0}$, it is called \textit{metric subspace}. Furthermore, it is said that $D_0$ is \textit{bounded} if there exists $M>0$ such that $d(x,y)\leq M$ for all $x,y\in D_0$. From now on we will assume that $(D,d)$ denotes a metric space and $(D_0,d\restriction_{D_0})$ a metric subspace of it. 
Recall that $(D,d)$ is a compact metric space if, and only if, for every sequence $\{a_n\}_n\subseteq D$ there exists $\{a_{n_k}\}_k$ a subsequence of $\{a_n\}_n$ that converges to some $a\in D$. This property is sometimes called the sequential characterisation of compactness in metric spaces.

The maps we are interested in in this paper are those that satisfy the Lipschitz condition. If $(D,d)$ and $(R,r)$ are metric spaces, a map $f\colon D\to R$ is called \textit{Lipschitz} if there exists a constant $L\in\R$ such that
\begin{equation*}
    r(f(x),f(y))\leq L d(x,y), \ \text{for all} \ x,y\in D.
\end{equation*}
It is said that $L$ is a \textit{Lipschitz constant} of $f$ and we say $f$ is $L$-Lipschitz to emphasize this constant. The infimum of all constants satisfying last inequality is the \textit{Lipschitz norm} of $f$, denoted by $Lip(f)$.  This can be written as $$Lip(f)=\underset{x\neq y}{\underset{x,y\in D}{\sup}}\frac{r(f(x),f(y))}{d(x,y)}.$$




Suppose $f$ is a real function defined on $D$, but known only in $S\subset D$. Several classical results concern the extension of $f$ to $D$ while preserving some initial properties of $f$ in $S$. The main result in this direction is when the property we want to preserve is the Lipschitz condition, and is called the McShane-Whitney theorem \cite{mcshane1934extension,whitney1992analytic}. Because of the useful extension formulae that can be obtained, this result has been the starting point for much research on the subject.

\begin{theorem}\label{teo:lipext}
If $f\colon S \subset D\to\R$ is an $L$-Lipschitz function, then there exists an $L$-Lipschitz function $F\colon D\to\R$ such that $F\restriction_{S}=f$.
This function $F$ is not unique, and two possible formulae are
$$
F^M(x):=\sup_{y\in S}\{f(y)-Ld(x,y)\} \quad \text{and} \quad F^W(x):=\inf_{y\in S}\{f(y)+Ld(x,y)\}, \quad x \in D,
$$
which are known as the McShane and Whitney extensions respectively.
\end{theorem}

\begin{remark}
We notice that any extension $F$ of a Lipschitz function $f$ verifies that
$F^M\leq F\leq F^W$. Moreover, for any $t\in (0,1)$, $F:=t F^W + (1-t) F^M$ is also a Lipschitz extension of $f$ with the same constant.
\end{remark}

The explanation and proofs  of general results on Lipschitz maps can be found in \cite{lipschitzbook}.

\subsection{Composition metrics and modulus of continuity} \label{sec:philipschitz}

We now focus on the possibility of generalising the Lipschitz condition to include a broader class of functions, satisfying a more relaxed criterion, and thereby preserving the extension theorems. This paper will address this in a positive way by introducing the concept of $\phi$-Lipschitz functions. First, we will define the class of functions $\Phi,$ which will serve as a form of generalised continuity module \cite{efi}. However, our main focus will be on metrics that can be formulated as compositions of a pre-existing metric with these functions, rather than on the functions themselves. Our specific interpretation of such a continuity module is outlined below:

\begin{definition}
    We will say that a function $\phi\colon\R^+\to\R^+$ belongs to $\Phi$  if for each $x,y\in\R^+$ it holds that
\renewcommand{\theenumi}{(\roman{enumi})}%
\begin{enumerate}
    \item $\phi(x+y)\leq \phi(x)+\phi(y) \ \text{(subadditivity)},$
    \item $\phi(x)<\phi(y) \ \text{if} \ x < y \ \text{(strictly monotonically increasing)},$
    \item $\phi(0)=0$,
    \item $\phi$ is continuous in $\R^+$.
\end{enumerate}
\hspace{\parindent} Let us remark that conditions (ii) and (iii) guarantee that $\phi(x)\geq0$ for all $x\in\R^+$, so these functions are well defined.
\end{definition}

As we said, such functions are often called modulus of continuity. A modulus of continuity is a non-decreasing function $\omega\colon[0,+\infty)\to[0,+\infty)$ that $\omega(0)=0$ and are continuous at $0$, but depending on the context they may satisfy other conditions. In fact, in \cite[Chapter~6, Section~4]{lipschitzbook} continuity modules are related to metrics in a similar way to the one we present here, although restricted to the context of the study of uniform continuity.


Now, following the scheme of the definition of the Lipschitz map, we will introduce a function $\phi\in\Phi$ of the distance. The definition is as follows.

\begin{definition}
Let $(D,d)$ and $(R,r)$ be metric spaces and let $f\colon D\to R$ be a map.  For $\phi\in\Phi$ we will say that $f$ is a $\phi$-\textit{Lipschitz map} if there exists $K>0$ such that
\begin{equation*}
    r(f(x),f(y))\leq K \phi(d(x,y)), \ \text{for all} \ x,y\in D.
\end{equation*}
As in the standard case, the constant $K>0$ is a Lipschitz constant of the $\phi-$Lipschitz map $f$. From now on, when we will compose the function $\phi\in\Phi$ with the metric $d$ we will write $d_{\phi}$. That is, $d_{\phi}=\phi\circ d$.
\end{definition}

We observe that every $L$-Lipschitz map is also a $\phi$-Lipschitz map for $\phi(x)=x$ and $K=L$. In addition, consider the function $f\colon\R^+\to\R^+$, defined as $f(x)=\sqrt{x}$. While this function is not Lipschitz, it meets the criteria of being $\phi$-Lipschitz for $\phi=f\in\Phi$ and $K=1$. Thus, this new category of mappings effectively generalises the original set of $L$-Lipschitz mappings. Moreover, the boundary of the distance between the images of two points given by the Lipschitz condition can be improved by considering it as $\phi$-Lipschitz. This case, which as we will see in Section \ref{sec:indexspaces} is fundamental in practical matters, is presented for example by the function $f(x)=\log(x+1)$. We know that $f$ is 1-Lipschitz and therefore $d(f(x),f(y))\leq d(x,y)$, but $f$ is also $\phi$-Lipschitz for $\phi=f$ and $K=1$, and therefore $d(f(x),f(y))\leq\log(d(x,y)+1)$, which is a better bound because $\log(d(x,y)+1)\leq d(x,y)$.

In fact, our proposal is to redefine the distance in such a way that the resulting metric space contains new maps that satisfy the proposed condition (in addition to the Lipschitz maps). In this way, by varying $\phi$, we have direct control over how to modify the original metric so that the resulting metric is better suited to the problem posed. Next we will see that $d_{\phi}$ does indeed define a distance.

\begin{proposition}\label{prop:dphimetric}
    Let $(D,d)$ be a metric space and $\phi\in\Phi$. Then $d_{\phi}$ is a metric.
    \begin{proof}
    Since $\phi$ is an injective function (because it is strictly monotonic) and $\phi(0)=0$, it is clear that $d_{\phi}(x,y)=0$ if, and only if, $d(x,y)=0$. And since $d$ is a metric this happens if, and only if, $x=y$, and we have just proved the identity of the indiscernibles for $d_{\phi}$. From the symmetry of $d$ as metric we directly deduce the symmetry of $d_{\phi}$. Taking into account the triangle inequality, the monotonicity and the subadditivity of $\phi$ we get $$d_{\phi}(x,z)\leq\phi(d(x,y)+d(y,x))\leq d_{\phi}(x,y)+d_{\phi}(y,z),$$ so $d_{\phi}$ satisfies the triangle inequality too.
    \end{proof}
\end{proposition}

The reciprocal of this result is usually false. For example, if $D=\R$ and we let $\phi\in\Phi$ be $\phi(x)=\sqrt{x}$ for $x\geq0$, and $d(x,y)=\abs{x-y}^2$, we know that $d_{\phi}=\abs{x-y}$ is a metric on $D$, but $d$ is not. To see this, it is enough to take $x=0$, $y=1$ and $z=3$ and check that $d(x,z)>d(x,y)+d(y,z)$, and therefore $d$ does not satisfy the triangular inequality. However, there is a kind of reverse implication. From the assumption that for any metric space $(D,d),$ the function $d_{\phi}$ is a metric, we can deduce some conditions that $\phi$ must satisfy. For example, it is easy to check that $\phi(0)=0$, and we will see next that $\phi$ must be subadditive. Let $x,y\in\R^+$ and define a metric space $D=\{a,b,c\}$ such that $d(a,b)=x, d(b,c)=y$ and $d(a,c)=x+y$.  Such a $d$ is clearly a metric in $D.$ As $d_{\phi}$ is a metric, by the triangle inequality we obtain $\phi(x+y)=d_{\phi}(a,c)\leq d_{\phi}(a,b)+d_{\phi}(b,c)=\phi(x)+\phi(y)$, and so $\phi$ is subadditive. Thus, \textit{ if a function $\phi: \mathbb R^+ \to \mathbb R^+$ satisfies that
 $d_\phi$ is a metric for all metrics $d,$ then $\phi$ is subbaditive and $\phi(0)=0$}. Nevertheless, it may happen that the conditions of monotonicity and continuity that $\phi$ muat satisfy are not met. That is, $d_{\phi}$ can be a metric for all metric space $(D,d)$ without $\phi$ being strictly monotonically increasing or continuous. For example, consider $\phi$ defined as $\phi(x)=0$ if $x=0$ and $\phi(x)=1$ if $x>0$. For each  metric space $(D,d)$ we note that
\begin{enumerate}
    \item For $a,b\in D$, condition $d_{\phi}(a,b)=0$ holds if, and only if, $d(a,b)=0$  because $\phi$ is only null at 0. Since $d$ is a metric we have that it holds if, and only if, $a=b$, so $d_{\phi}$ satisfy the identity of indiscernibles.

    \item Since $d(a,b)=d(b,a)$ for all $a,b\in D$ it is clear that $d_{\phi}(a,b)=d_{\phi}(b,a)$, that is, $d_{\phi}$ is symmetric.

    \item Take $a,b,c\in D$ and write $x=d(a,b)$, $y=d(b,c)$ and $z=d(a,c)$. If $a,b,c$ are different from each other then $x,y,z>0$ since $d$ is a metric, and so $\phi(x)=\phi(y)=\phi(z)=1$. From this we conclude that $d_{\phi}$ satisfy the triangular inequality for $a,b,c$. In case some of the elements of $D$ that we have taken coincide, the triangular inequality of $d_{\phi}$ for them is trivial.
\end{enumerate}

\begin{center}
\begin{figure}[h]
    \captionsetup{justification=centering,margin=-2cm}
    \hspace{-4em}\includegraphics[scale=0.8]{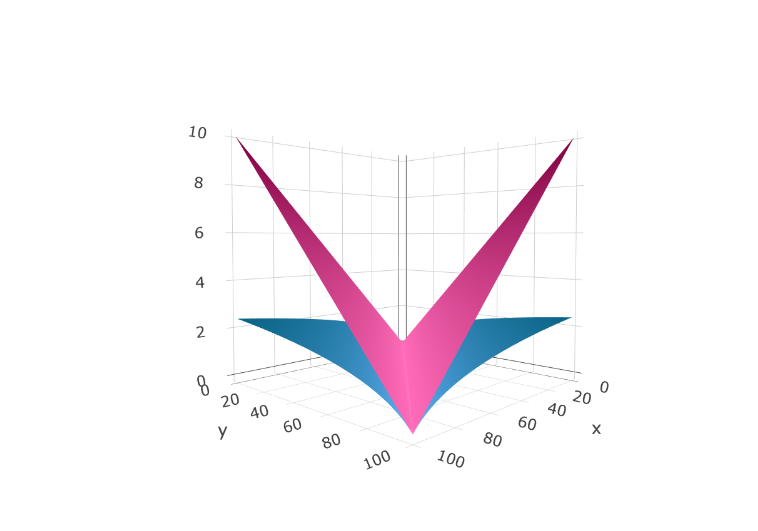}
    \caption{Comparative between $d(x,y)$, in pink, and $d_{\phi}(x,y)$, in blue.}
    \label{fig:ddphi}
\end{figure}
\end{center}

\begin{center}
\begin{figure}[h!]
\centering
\begin{subfigure}{.45\textwidth}
  \centering
  \begin{varwidth}{\linewidth}
  \hspace{-10em}\includegraphics[scale=0.5]{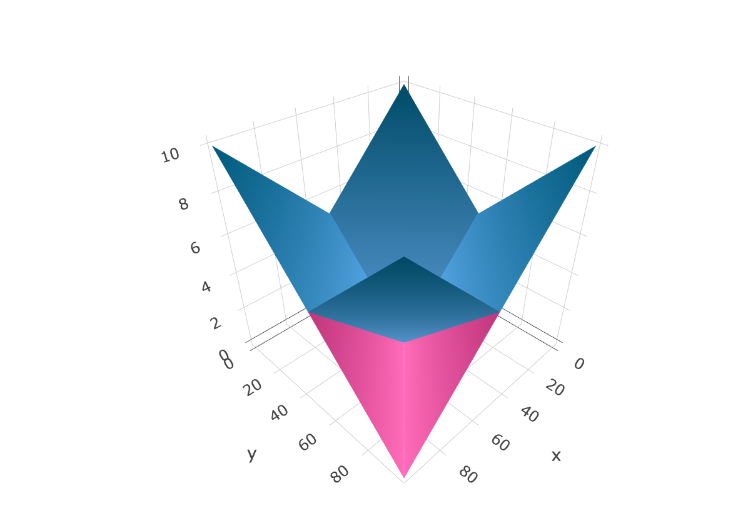}
  \caption{Comparison between $d(x,y)$, in pink, and $d(x,0)+d(0,y)$, in blue.}
  \end{varwidth}    
  \label{fig:dtd}
\end{subfigure}
\begin{subfigure}{.45\textwidth}
  \centering
  \begin{varwidth}{\linewidth}
  \hspace{-7em}\includegraphics[scale=0.5]{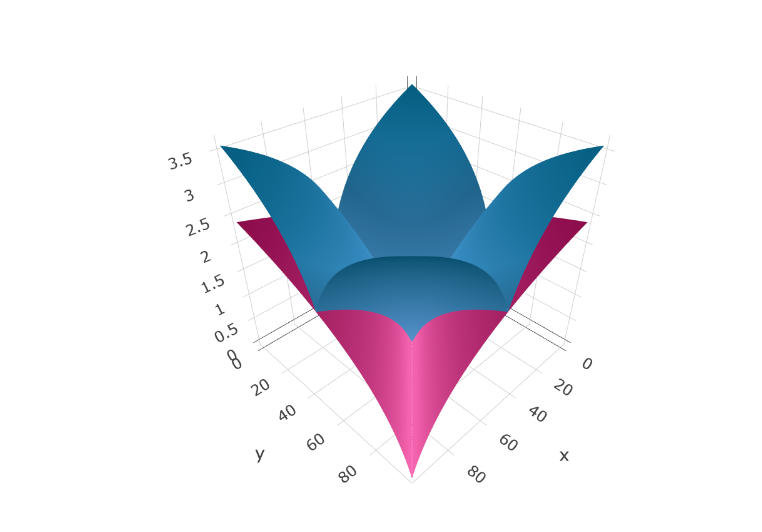}
  \caption{Comparison between $d_{\phi}(x,y)$, in pink, and $d_{\phi}(x,0)+d_{\phi}(0,y)$, in blue.}
  \end{varwidth} 
  \label{fig:dtdphi}
\end{subfigure}
\caption{\small Representation of the triangular inequality of $d$ and $d_{\phi}$. \normalsize}
\label{fig:test}
\end{figure}
\end{center}

For a better understanding of how $d$ and $d_{\phi}$ relate to each other, we present in Figures \ref{fig:ddphi} and \ref{fig:test} a graphical representation of the behavior of two concrete metrics: the euclidean metric $d$ in $\R$ and its composition with $\phi(x)=\log(1+x)$, that is, $d_{\phi}$. In the Figure \ref{fig:ddphi} it can be seen that $d_{\phi}$ "smooths" the distance between two real numbers with respect to $d$. More specifically, $d$ and $d_{\phi}$ have a similar behavior for $x$ and $y$ "close" to each other, but for more distant values $d_{\phi}$ attenuates growth with respect to $d$. Figure \ref{fig:test} provides a comparison between the behavior of the triangular inequality of $d$ and $d_{\phi}$. As before, we can see how the logarithmic growth of ${\phi}$ carries over to $d_{\phi}$, and also to its triangular inequality.

\newpage

\begin{example}
More examples of $\phi$-Lipschitz maps are presented below.
    \begin{itemize}
        \item A map $f\colon \R^n\to \R$ is called $\alpha$-\textit{Hölder continuous} if there exist $C>0$ and $\alpha>0$ such that
        $$
        r(f(x),f(y))\leq Cd(x,y)^{\alpha}, \quad \text{for all} \ x,y\in D.
        $$
        These maps are $\phi$-Lipschitz maps for $K=C$. Indeed, if $\alpha\in(0,1]$ then $\phi$ defined as $\phi(x)=x^{\alpha}$ belongs to $\Phi$ and $f$ is $\phi$-Lipschitz for $K=C$. If $\alpha>1$ it can be proven that $f$ is a constant map, so it is $\phi$-Lipschitz for any $\alpha\geq0$.

        \item Consider $\R$ equipped with its usual metric and let $\phi\in\Phi$. From the subadditivity it follows that $\phi(x)-\phi(y)\leq\phi(x-y)$ for $x\geq y$, so $d(\phi(x),\phi(y))\leq\phi(d(x,y))$. That is, every $\phi\in\Phi$ is a $\phi$-Lipschitz function with $K=1$. Examples of $\phi$ functions, apart from those already presented, are $\phi(x)=\arctan(x)$, $\phi(x)=x(x+1)^{-1}$ or $\phi(x)=x(x^2+1)^{-1/2}$.
    \end{itemize}
\end{example}

Compactness for subsets of the metric spaces $(D,d_\phi)$ in terms of compactness of $(D,d)$ is straightforward. 

\begin{proposition}\label{prop:dphicompacto}
Let $\phi \in \Phi.$
    If $(D,d)$ is compact, then $(D,d_{\phi})$ is compact too.
    \end{proposition}

\section{Index extension}\label{sec:indexspaces}

The aim of this section is to adapt the concepts of index and index space in the terms that were introduced in \cite{index} to the context of the composition metrics. We will also explain the extension methods available to obtain an approximation framework for Lipschitz regression. For this purpose, we will present two different techniques. The first is based on what are known in \cite{index} as standard indices, which are essentially defined by the distance to a reference point. The second is based on the McShane and Whitney extension formulae already explained.

Our models essentially consist of a metric space $(D,d)$ and an index $I$ (a real Lipschitz function) defined on it. These indices give meaningful values to the elements of $D$, which are the objects of the models. The Lipschitz condition of the index is the tool that introduces a certain concordance of the metric with the nature of the distance in $D$. Thus, for example, a ranking of the elements of $D$ can be constructed according to the value of their index. The essence of the research we propose in this paper is that we introduce another tool to improve models based on index spaces: the construction of composition metrics $d_\phi$ that improve the fit of the metric to the situation we want to model.

In \cite{index} it is said that a triplet $(D,d,I)$ is an index space.
The index $I\colon D\to\R$ is bounded if $\textstyle\sup_{a\in D}\abs{I(a)}\leq C$ for some $C>0.$  The infimum of all such constants $C$ satisfying the last inequality is denoted by $B(I)$. That is, $B(I)=\sup_{a\in D}\abs{I(a)}$. In case we are considering the normed space structure of the space of functions with the uniform norm, we will write $\| \cdot\|_\infty$ instead of $B(\cdot).$

We will need two more boundedness type properties to characterize the set of indexes we are interested in.

\begin{definition}
    Let $(D,d)$ be a metric space. An index $I\colon D\to\R$ is $Q$-\textit{normalized}, for $Q>0$, if
    $$d_{\phi}(a,b)\leq Q(\abs{I(a)}+\abs{I(b)}), \quad \text{for all} \ a,b\in D.$$
These functions are often called  \textit{Katetov functions}. The infimum of all such constants $Q$  is the \textit{normalization constant} of $I$; we write  $N(I)$ for it. 
\end{definition}
%
%
\begin{definition}
    Let $(D,d)$ be a metric space. An index $I\colon D\to\R$ is $\phi$-\textit{coherent} for $\phi\in\Phi$ and a certain constant $K>0$ if
    $$\abs{I(a)-I(b)}\leq K d_{\phi}(a,b), \quad \text{for all} \ a,b\in D,$$
    that is,  if the index $I$ is $\phi$-Lipschitz for the constant $K$. The infimum of all constants $K$ satisfying last inequality is the \textit{coherence constant} of $I$, and we will write for it $C(I)$.
\end{definition}


In practice, when building models using index spaces, the index value will not always be available for all elements of the metric space. This may be the case, for example, when it is costly or impossible to obtain all the data for the situation on which the model is built. In this context, it is useful to have tools that allow us to approximate the index from non-indexed elements. As mentioned above, in this section we propose two ways of doing this: the first is to identify/approximate the index of interest using what we will call standard indices, while the second is to approach the question as a Lipschitz regression problem using the classical extension formulae. Both techniques can be adapted to the case of composition metrics, as we will see below.

\subsection{Approximation through standard indices}

In a metric space, the distance function $d$ allows us to define indices that are inherently natural, known as standard indices. For a given point $a \in D$, we can define an index $I_a$ as $I_a(b):=d(a,b),$ where $b \in D$. These functions are called standard indices. Typically, the selection criterion for $a$ is to identify the element that minimises certain properties, although any other element could be chosen. The aim of this section is to introduce an additional tool, the composition metric $d_\phi$, which allows us to use the associated standard indices $b \mapsto d_{\phi}(a,b)$. Our interest, therefore, is to show how we can approximate any given index using standard indices. We will first present some results on compactness in index spaces in order to establish the necessary theoretical framework. We will then turn our attention to the central issue of approximation. The main goal is to show that a robust approximation method is possible, at least in theory. 

Note that uniformly bounded sets of indices (without Lipschitz-type requirements) are pointwise compact. Indeed, let $C>0$ and consider the set of functions $\mathcal{F}_C:=\{I:D\to\R:B(I)\leq C\}$. In it we can define two natural topologies: the \textit{uniform topology} (for the norm $B(I):= \sup_{a \in D} |I(a)|$) and the \textit{topology of pointwise convergence}. 
For $ -C \le k \le C$ consider $\mathcal{F}_{C,k} :=\{I:D\to\R: k \le B(I)\leq C\}$, it is a \textit{compact space with the topology of pointwise convergence}, since we can identify each function of $\{I_{\eta}\}_{\eta\in\Lambda}$ with an element of $\Pi_{a\in D}[k,C]$ (product of compact spaces in the product topology) and by Tychonoff's theorem gives that it is compact.

However, although we cannot expect compactness with respect to the topology of uniform convergence of the entire space of uniformly bounded functions, the space of standard indices is compact.

\begin{proposition}\label{cor:compwc}
Let $\phi \in \Phi.$ Let $(D,d)$ be a compact metric space. Then there exists $C>0$ such that the space $S:=\{d_{\phi}(a,\cdot):a\in D\}$ of standard indices associated with the metric $d_\phi$, is included in  $\mathcal{F}_C$ and it is compact with respect to the topology of uniform convergence.
    \begin{proof}
    First, note that due to the compactness of $(D,d)$ the metric $d$ is bounded by a certain constant $R>0$. Then $S$ is bounded by $C:=\phi(R)$ and so $S\subset \mathcal{F}_C$. In addition, let $\{d_{\phi}(a_n,\cdot)\}_n\subset S$. By the compactness of $(D,d)$ again and by Proposition \ref{prop:dphicompacto} we know that $(D,d_{\phi})$ is also compact, so the sequence $\{a_n\}_n$ admits a subsequence $\{a_{n_k}\}_k$ that  converge to $a_0\in D$ with respect to $d_{\phi}$. So, for every $\varepsilon>0$ there exists $k_0\in\mathbb{N}$ such that $d_{\phi}(a_{n_k},a_0)<\varepsilon$ for all $k\geq k_0$. Then, for every $b \in D,$ by the triangular inequality we get
$$
\sup_{b\in D}\abs{d_{\phi}(a_{n_k},b)-d_{\phi}(a_0,b)}\leq \sup_{b\in D}d_{\phi}(a_{n_k},a_0)=d_{\phi}(a_{n_k},a_0)<\varepsilon,
$$
    for $k\geq k_0$. That is, $\{d_{\phi}(a_{n_k},\cdot)\}_k$ converges uniformly to $d_{\phi}(a_0,\cdot)$, so by the sequential characterization of compactness we conclude the desired result. \end{proof}
\end{proposition}

This allows us to obtain that there is always a best approximation for each Lipschitz index $I$ by the elements of $S,$ as stated in Corollary \ref{besta}.

\begin{corollary} \label{besta}
Let $\phi \in \Phi$ and let $(D,d)$ be a compact metric space.
For every Lipschitz index $I$ acting in the compact space $(D,d_\phi),$ there is an element $a_0 \in S$ such that the standard index $I_{a_0,\phi}$ is the best approximation to $I$ in $S,$ that is,
$$
\inf \{ \| I - I_{a,\phi} \|_\infty :a \in D\} = \| I - I_{a_0,\phi} \|_\infty.
$$

\begin{proof} A direct compactness argument gives the proof. The function $F: S \to \mathbb R$ defined as
$$
F(I_{a,\phi}) = \| I- I_{a,\phi}\|_\infty,  \quad a \in D,
$$
is continuous (and even Lipschitz), since
$$
\big| F(I_{a,\phi})  - F(I_{b,\phi}) \big| = \big| \| I- I_{a,\phi}\|_\infty - \| I- I_{b,\phi}\|_\infty \big| 
\le \big\| I_{a,\phi} - I_{b,\phi}\|_\infty 
$$
$$
= \sup_{c \in D}  \big| \phi \circ d(a,c)- \phi \circ d(b,c) \big|
\le \phi \circ d(a,b) = d_\phi(a,b),
$$
for all $a,b \in D.$ Compactness of $S$ (Proposition \ref{cor:compwc}) proves the result.
\end{proof}
\end{corollary}


Let us now focus our attention on the approximation bound for general Lipschitz indices. In particular, we will show that we can always get an estimate of the error made by this approximation procedure in terms of the generalisation of the normalisation and the coherence constants.  For technical reasons, we also need to name a type of sequence that will appear in the result we are looking for. 

A sequence $\{a_n\}_n\subset D$ is \textit{pointwise Cauchy} if for each $b\in D$ there exists $\lim_n d(a_n,b)$. Every convergent sequence is pointwise Cauchy, but the converse is false. The sequence $a_n=1/n$ in $D=(0,1]$ with the Euclidean metric is an  example of non-convergent pointwise Cauchy sequence. The definition above can directly be translated for composition metrics $d_\phi,$ $\phi\in\Phi$. 

The following result will finally give us the tool to approximate an index using the standard indices, in a sense of improvement of \cite[Th. 1]{index}. We will also see that this approximation depends on the product of the normalisation and coherence constants. Specifically, we will see these results for the following set of indices to be approximated:
$$
\mathcal{R}_{K,Q,C}:=\left\{ I\geq0:
\begin{matrix}
\abs{I(a)-I(b)}\leq Kd_{\phi}(a,b),  \\
\tfrac{1+KQ}{K}\inf(I)+d_{\phi}(a,b)\leq Q(I(a)+I(b)), \\
B(I)\leq C 
\end{matrix} 
\right\}.
$$

The proof of the following result is similar to the theorem discussed before. We include the proof for the aim of completeness.

\begin{theorem}\label{teo:apie}
Let $K,Q >0$ such that $KQ \geq 1.$
For every $I\in\mathcal{R}_{K,Q,C}$
there exists a pointwise Cauchy sequence $\{a_n\}_n$ such that $I(b)\leq\inf(I)+\lim\limits_n Kd_{\phi}(a_n,b)\leq KQI(b)$, for each $b\in D$.

\begin{proof}
    Take $b\in D.$  Fix $n\in\mathbb{N}$. Then for every $n \in\mathbb{N}$ there is an element  $a_n\in D$ such that $I(a_n)-\tfrac{1}{n}\leq\inf(I)$ and so
    \begin{align*}
        \inf(I)+Kd_{\phi}(a_n,b)&\leq KQI(a_n)+KQI(b)-KQ\inf(I)\\
        &\leq KQI(b)+KQI(a_n)-KQ\left(I(a_n)-\tfrac{1}{n}\right)=KQI(b)+\tfrac{KQ}{n}.
    \end{align*}
    In addition,
    $$I(b)-I(a_n)\leq\abs{I(b)-I(a_n)}\leq Kd_{\phi}(a_n,b),$$
    and therefore
    \begin{equation*}
        I(b)\leq Kd_{\phi}(a_n,b)+I(a_n)\leq Kd_{\phi}(a_n,b)+\inf(I)+\tfrac{1}{n}\leq KQI(b)+\tfrac{1+KQ}{n},
    \end{equation*}
    for all $n\in\mathbb{N}$ and $b\in D$. Now note that $d_{\phi}(a_n,\cdot)=I_{a_n}(\cdot)\in S$, thus, from the compactness of $S$ seen in Proposition \ref{cor:compwc} there exists a subsequence $\{a_{n_k}\}_k$ such that $\lim\limits_kd_{\phi}(a_{n_k},b)=d_{\phi}(a_{0},b)$ for each $b\in D$ and certain $a_0\in D$. Therefore $\{a_{n_k}\}_k$ is a pointwise Cauchy sequence,  and from the last inequality  we conclude that
    $$I(b)\leq \inf(I)+K\lim\limits_kd_{\phi}(a_{n_k},b)\leq KQ I(b)$$
    for every $b\in D$.
\end{proof}
\end{theorem}

This result leads us to identify the function $\tilde{I}:=\inf(I)+K\lim\limits_kd_{\phi}(a_{n_k},\cdot)$ as an approximation for $I\in\mathcal{R}_{K,Q,C}$. The maximum error committed is bounded by
$$
\sup_{b\in D}\abs{\tilde{I}(b)-I(b)}\leq\sup_{b\in D}\abs{KQI(b)-I(b)}=(KQ-1)C.
$$
So if $KQ\approx 1$ the comparison between $\tilde{I}$ and $I$ is reasonable, but as these constants increase the approximation may get worse. 

In addition, taking $b=a_0$ in the result of Theorem \ref{teo:apie} we have $$I(a_0)\leq \inf(I)+K\lim\limits_kd_{\phi}(a_{n_k},a_0)=\inf(I)+Kd_{\phi}(a_0,a_0)=\inf(I),$$ and so $I(a_0)=\inf(I)$. Consequently we can write $\Tilde{I}(\cdot)=I(a_0)+d_{\phi}(a_0,\cdot)$, being $a_0\in D$ a point in which $I$ attains their minimum. In addition, we can assume for the indices we work with that $\inf(I)=0$, so we can make this approximation for any $Q$-normalised and $\phi$-coherent index for $K$. In this case $K\tilde{I}(\cdot)=d_{\phi}(a_0,\cdot)$.

\begin{remark}
The condition $QK\geq 1$ stated in Theorem \ref{teo:apie} is in a sense universal in index spaces. Indeed, since $I\geq0$ and $I$ attains the minimum in $b\in D$ (a circumstance that arises, for instance, when $D$ is finite or compact) and this minimum is 0, then
\begin{align*}
    I(a)&=I(a)-I(b)=\abs{I(a)-I(b)}\leq Kd_{\phi}(a,b)\\
    &\leq KQ(I(a)+I(b))=KQI(a).
\end{align*}
Consequently, if there exists a point $b\in D$ such that $I(b)=0$, we can ensure $KQ\geq 1$. Moreover, even if such a scenario does not arise, in the case of an index defined within a compact metric space, we will still have $\inf(I)=I(b)$ for some $b\in D$, implying that $I_0(a):=I(a)-I(b)$ constitutes another positive index that retains the same order properties as $I$, with $I_0(b)=0$.
\end{remark}

\subsection{Extension theorems: McShane and Whitney formulas}

Another approach to the extension problem is to formulate an approximation for the unknown values using the extension formulas outlined in section \ref{sec:basic}. While there are alternative expansion techniques, such as those discussed in \cite{absolute}, which may be more suitable for specific purposes, for our overarching application it proves more advantageous to use the classical McShane and Whitney formulas. The method follows the procedure explained in \cite{index}, adapted to the case of the composition metric. A full explanation can be found in Section 4 of that article, and in the example given in Section 5 of the same article. 

  \begin{remark}
    To ensure that the composition $d_\phi$ defines a metric, in the proof of Proposition \ref{prop:dphimetric} we only considered properties (i)-(iii) of $\Phi$. That is, the continuity property is not necessary, although it was needed to ensure the compactness results for standard indices and to obtain the result of the Theorem \ref{teo:apie}. However, this is not the case for these results on the McShane-Whitney extensions, since all that is required is that $d_\phi$ defines a metric. Thus, the functions of $\Phi$ can be more general than a continuity modulus for this technique.
\end{remark}

Convex combinations of the classical extension formulas can also be used. Furthermore, as in the case of standard indices, the product of the normalisation and coherence constants is related to the accuracy of the approximation. Note that
\begin{align*}
    I^W(a) &= \inf_{b\in D} \{I(b)+K d_{\phi}(a,b)\}\\
    &\leq \inf_{b\in D} \{I(b)+KQ(I(a)+I(b))\}\\
    & = (1+KQ)\inf_{b\in D} I(b) + KQI(a),
\end{align*}
for any $a\in D$. So,
$$
\sup_{a\in D} \abs{I^W(a)-I(a)} \leq (1+KQ)\abs{\inf_{b\in D} I(b)} + \abs{KQ-1}\sup_{a\in D} \abs{I(a)}.
$$
If we suppose that $\inf_{b\in D} I(b)=0$ and $KQ\geq1$, as in the case of standard indices, the last bound is reduced to 
$$
\sup_{a\in D} \abs{I^W(a)-I(a)}\leq (KQ-1)C,
$$
 what indicates that the approximation improves when $KQ \to 1.$ Similar bounds can be found for the McShane formula, and then for any convex combination of both.

\section{Applications: the liveability index for cities} \label{sec:apliccation}

     In this section we present a methodological proposal for implementing the theoretical content of this paper in order to extend a given index. We start with a finite set of elements, characterised by certain real variables, for which we want to determine the value of a certain index $I$. For some of these elements, the value of the index of interest is already known. Our aim is to extrapolate this known information to estimate the value of the index for those elements where it is not defined. Mathematically, this set constitutes a metric space $D$, equipped with an appropriate distance metric related to the nature of the data or the problem at hand. In addition, there is a known index within a subset $D_0\subset D$ that we want to extend to the whole space $D$. To achieve this goal, we propose the following methodology.

     \subsection{Methodology}\label{subsec:method}
The first question we need to address is whether the diverse nature of the variables could distort the metric under consideration, given the heterogeneity of their scales. To mitigate this potential problem, we suggest normalising the variables to a common scale by subtracting the minimum value and dividing by the range. More precisely, let $D=\{y_j\}_{j=1}^n$ and $y_j=(x_1^j,\cdots, x_m^j)$. Let $a_k:=\max_j x_k^j$ and $b_k:=\min_j x_k^j$ for each $k=1,\cdots,m$. We then transform $$x_k^j \sim \frac{x_k^j-b_k}{a_k-b_k},$$ for all $j$ and $k$, so that we have the new variables restricted to [0,1], in the same scale.

To assess the accuracy of the approximation we are about to make, we need a measure of the error made, which in our case will be the Root Mean Square Error (RMSE). This gives the expected absolute error and is defined as
$$\text{RMSE} =\sqrt{\frac{1}{n} \sum_{j=1}^n\left(\Tilde{I}(a_j)-I(a_j)\right)^2},$$
    where $a_1,\ldots,a_n$ are the observations where we want to estimate the error, and $\Tilde{I}$ is the approximation to $I$. However, since we have no information about the index values we want to approximate, we need a strategy to estimate this error. In our approach, we will divide $D_0$, the subset of observations with known indices, into two subsets: a training set and a test set. We will use observations from the training set, which is seventy percent of the total observations and is randomly selected, to perform the expansion. We will then use the remaining observations from the test set to calculate the Root Mean Square Error (RMSE). Nevertheless, the randomness inherent in this process can affect the resulting error, so it may not be representative. To mitigate this problem, we use a technique known as cross-validation. This involves repeating the process multiple times (twenty times in our case) in order to accumulate the resulting errors and providing a more robust conclusion regarding accuracy.

In order to select a $\phi$ function that best fits the model, we engage in an optimisation process aimed at minimising the error bound on the test set. According to theorem \ref{teo:apie}, this means minimising the product of the coherence and normalisation constants. Ideally, we would divide our dataset into three subsets: the training and test subsets mentioned above, and a validation subset from which we would perform the fitting. However, due to the potentially limited number of observations available, the resulting subsets may not be sufficiently significant for this study. Therefore, we will use the values obtained from the test set as a reference for our optimisation process. To do this, we will consider that the linear combination of functions in $\Phi$ with positive scalars is another function in $\Phi$. We will first choose a set of elementary functions $\{\phi_j\}_{j=1}^n\subset\Phi$ and we will discuss for which values $\lambda_1,\ldots,\lambda_n\geq 0$ the function $\phi:=\lambda_1\phi_1+\ldots+\lambda_n\phi_n$ ensures that the metric $d_{\phi}$ is optimal in terms of the bound. To do this, we will consider the particle swarm optimisation algorithm of the $R$ library "pso". This type of algorithm, unlike those based on the gradient of the function, explores the entire possible set of parameters and thus avoids convergence to local minima.

If we consider the extension using the Whitney and McShane formulae, which are maximum and minimum extensions respectively, we can ask whether we can consider an intermediate extension that minimises the error. That is, for which $\alpha\in[0,1]$ the extension $I:=(1-\alpha)I^W+\alpha I^M$ minimises the error. As in the previous point, the preferred way to obtain this parameter would be from a validation set, but for the reasons already explained, we will use the test set as a reference. To do this, we will choose the value of $\alpha$ according to the following result.

    \begin{proposition}
        Let $(D,d)$ be a finite metric space and $I\colon D_0\subset D\to\R$ a $\phi$-coherent index for $K>0$. Let $S_1,S_2\subset D_0$ such that $S_1\cup S_2=D_0$ and $S_1\cap S_2 = \emptyset$. Consider
        $$I^W(b):=\inf_{a\in S_1}\{I(a)+Kd_{\phi}(a,b)\}, \quad I^M(b):=\sup_{a\in S_1}\{I(a)-Kd_{\phi}(a,b)\}.$$
        Naming $I^E_{\alpha}:= (1-\alpha)I^W+\alpha I^M$, then 
$$
\min_{0\leq\alpha\leq 1}\sum_{b\in S_2}\left(I(b)-I^E_{\alpha}(b)\right)^2=\sum_{b\in S_2}\left(I(b)-I^E_{\alpha_0}(b)\right)^2,
$$ 
for
 $$
\alpha_0=\frac{\sum_{b\in S_2}\left(I^W(b)-I(b)\right)\left(I^W(b)-I^M(b)\right)}{\sum_{b\in S_2}\left(I^W(b)-I^M(b)\right)^2}.
$$

        \begin{proof}
            Let $F(\alpha):=\sum_{b\in S_2}\left(I(b)-I^E_{\alpha}(b)\right)^2$ and note that
            $$F(\alpha)=\sum_{b\in S_2}\left(I(b)-I^W(b)+\alpha \left( I^W(b)-I^M(b)\right)\right)^2,$$
            so
            \begin{align*}
                F'(\alpha)&=2\sum_{b\in S_2}\left(I(b)-I^W(b)+\alpha \left( I^W(b)-I^M(b)\right)\right)\left( I^W(b)-I^M(b)\right)\\
                &=2\sum_{b\in S_2}\left(I(b)-I^W(b)\right)\left( I^W(b)-I^M(b)\right)\\
                &+2\alpha\sum_{b\in S_2}\left( I^W(b)-I^M(b)\right)^2.
            \end{align*}
            Solving the equation $F'(\alpha)=0$ we obtain the $\alpha_0$ we are looking for, since $$F''(\alpha)=2\sum_{b\in S_2}\left( I^W(b)-I^M(b)\right)^2\geq0.$$
        \end{proof}
    \end{proposition}
    We have chosen to use the RMSE to quantify the error instead of other alternatives because it ensures the differentiability of the function $F$ under consideration. This facilitates a straightforward analysis of its minimum, as demonstrated in the previous proposition. Had we chosen other error measures defined in terms of absolute value, such as Mean Absolute Error (MAE) or Symmetric Mean Absolute Percentage Error (SMAPE), it would have been necessary to use a more complicated and indirect approach to determine the optimal $\alpha$ due to the non-differentiable nature of such definitions.

If we consider the extension by identifying the index with a standard one, we will proceed to find the element $a_0\in D$ that minimises the index. In this case we will take $\Tilde{I}(b):= Kd_{\phi}(a_0,b)$ as an approximation of $I(b)$, since $\min(I)=0$ after scaling.

\subsection{Extension of a liveability index}
In the following, we will validate our proposed approach using the so-called \textbf{AARP Livability Index}. In 2018, 55 per cent of the world's population lived in urban areas, a figure that is expected to rise to 68 per cent by 2050, according to \cite{urban}. This rapid urbanisation underlines the growing importance of studying and quantifying concepts such as quality of life or liveability in cities, as evidenced by various indices outlined in \cite{liveabindex}. The purpose of these indices is twofold: first, to define the concept of livability and identify the parameters that describe it; and second, to provide insights into which cities or neighbourhoods offer better living conditions. Armed with this information, urban planners can identify areas in need of intervention, while public administrations can target development and investment initiatives to regions with poor living conditions. However, assessing liveability can be a complex task due to the large number of factors involved, many of which are difficult to assess. For example, The Economist's Global Liveability Index, one of the most widely used, includes thirty indicators across five categories, with some indicators building on others. In addition, certain factors, such as climate discomfort for travellers or levels of corruption, are subjective and inherently difficult to quantify. In this section, we propose to use the index extension theory developed in the previous section to approximate livability using only indicators related to alternative mobility. The aim is to shift the focus away from subjective or complex social indicators and instead focus on factors that can be easily estimated based on existing infrastructure and the connectivity of urban patterns.

Let us explain the databases we will consider. Walk Score$^{\text{\textregistered}}$\footnote{https://www.walkscore.com} is a website that scores, from 0 to 100, the walkability performance of 123 cities in the United States and Canada according to a series of parameters, such as intersection density, block length, access to amenities in less than a 5-minute walk, etc. In addition, Walk Score uses a similar scoring system to rate the performance of cities for walking and cycling. Our aim will be to use these three indicators to approximate the AARP livability index \footnote{https://livabilityindex.aarp.org/scoring}. This index evaluates 61 different indicators in seven categories (housing, neighbourhood, transportation, environment, health, engagement and opportunity) to assess the livability of US cities, such as housing costs, crime rates, air quality or income inequality. The final score is a number between 0 and 100, with 50 being an average score and higher scores corresponding to above-average performance and vice versa. In mathematical terms our metric space is $D\subset[0,100]^3$, where each element $(x,y,z)\in D$ represents a city with walk, transport and bike scores $x,y$ and $z$ respectively, equipped with the canonical metric $d$ of $\R^3$. For 101 US cities we have defined the index of interest, which we will call $I$, and our goal is to be able to define it for 22 Canadian cities as well. Table \ref{tab:example} shows an example of our data.

\begin{table}[h!]
    \centering
    \begin{tabular}{|c||ccc|c|}
    \hline
         City & Walk Score & Transit Score & Bike Score & $I$ \\
         \hline
         \hline
         New York & 88 & 88.6 & 69.3 & 63 \\
         Los Angeles & 68.6 & 52.9 & 58.7 & 49 \\
         Chicago & 77.2 & 65 & 72.2 & 57 \\
         Toronto & 61 & 78.2 & 61 & ? \\
         Houston & 47.5 & 36.2 & 48.6 & 48 \\
         Montreal & 65.4 & 67 & 72.6 & ? \\
         \hline
    \end{tabular}
    \caption{Examples of scores and index for some cities.}
    \label{tab:example}
\end{table}

We have implemented the extension of our index according to the considerations of \ref{subsec:method}. In particular, for illustrative purposes, we study our method by considering two linear combinations of functions of $\Phi$ as follows:
$$\phi(x)=p_1 x + p_2\log(1+x) + p_3 \arctan(x) + p_4 \frac{x}{1+x}, \quad p_1,\ldots, p_4\geq 0,$$
and
$$\psi(x)=p_1 \sqrt{x} + p_2\log\left(1+\sqrt{x}\right) + p_3 \arctan\left(\sqrt{x}\right) + p_4 \frac{\sqrt{x}}{1+\sqrt{x}}, \quad p_1,\ldots, p_4\geq 0.$$

Table \ref{tab:ciucompphi} presents a comparative analysis of the performance of the two extension procedures studied (identification with a standard index and McShane-Whitney extensions) and of the various metrics considered. The results show that the identification of our index with a standard index gives poor performance, as evidenced by a significantly high estimated mean error, making it unsuitable for consideration. As far as the extension formulas are concerned, our technique maintains the standard deviation of the error while reducing the expected RMSE, although not significantly. However, it should be noted that our technique requires a longer computation time, mainly due to the optimisation process involved. Furthermore, table \ref{tab:ciucomppsi} shows the results of considering a linear combination as $\psi$ in this scenario, alongside the results of the classical technique. In particular, there is a more significant improvement in performance compared to the original metric, particularly noticeable in the case of standard indices. This variance in performance highlights the fact that the choice of $\phi$-metric has a significant impact on the degree of improvement obtained relative to the original metric.

\begin{table}[h!]
    \centering
    \begin{tabular}{|c|c|c|c|c|}
    \hline
    \multirow{2}{*}{ Function $\phi$ } &\multicolumn{2}{|c|}{Standard} & \multicolumn{2}{|c|}{McShane-Whitney}\\
    \hhline{~----}
         & Lipschitz & $\phi$-Lipschitz & Lipschitz & $\phi$-Lipschitz\\
        \hline
        \hline
        Mean RMSE & 138.43 & 79.48 & 5.08 & 5.04 \\
        \hline
        Median RMSE & 140.49 & 81.25 & 5.12 & 5.05\\
        \hline
        Standard deviation & 24.11 & 8.78 & 0.61 & 0.60\\
        \hline
        Seconds per iteration & 1.039$\times 10^{-3}$ & 3.316$\times 10^{-1}$& 1.513$\times 10^{-3}$ & 3.330$\times 10^{-1}$\\
        \hline
    \end{tabular}
    \caption{Comparison of method performance for function $\phi$.}
    \label{tab:ciucompphi}
\end{table}

\begin{table}[h!]
    \centering
    \begin{tabular}{|c|c|c|c|c|}
    \hline
    \multirow{2}{*}{ Function $\psi$ } &\multicolumn{2}{|c|}{Standard} & \multicolumn{2}{|c|}{McShane-Whitney}\\
    \hhline{~----}
         & Lipschitz & $\psi$-Lipschitz & Lipschitz & $\psi$-Lipschitz  \\
        \hline
        \hline
        Mean RMSE & 138.43 & 16.69 & 5.08 & 4.55\\
        \hline
        Median RMSE & 140.49 & 16.50 & 5.12 & 4.47\\
        \hline
        Standard deviation & 24.11 & 1.52 & 0.61 & 0.63\\
        \hline
        Seconds per iteration & 1.039$\times 10^{-3}$ & 3.000$\times 10^{-1}$ & 1.513$\times 10^{-3}$ & 3.013$\times 10^{-1}$\\
        \hline
    \end{tabular}
    \caption{Comparison of method performance for function $\psi$.}
    \label{tab:ciucomppsi}
\end{table}

\begin{figure}[h!]
    \begin{subfigure}{0.485\textwidth}
        \centering
        \includegraphics[width=\linewidth]{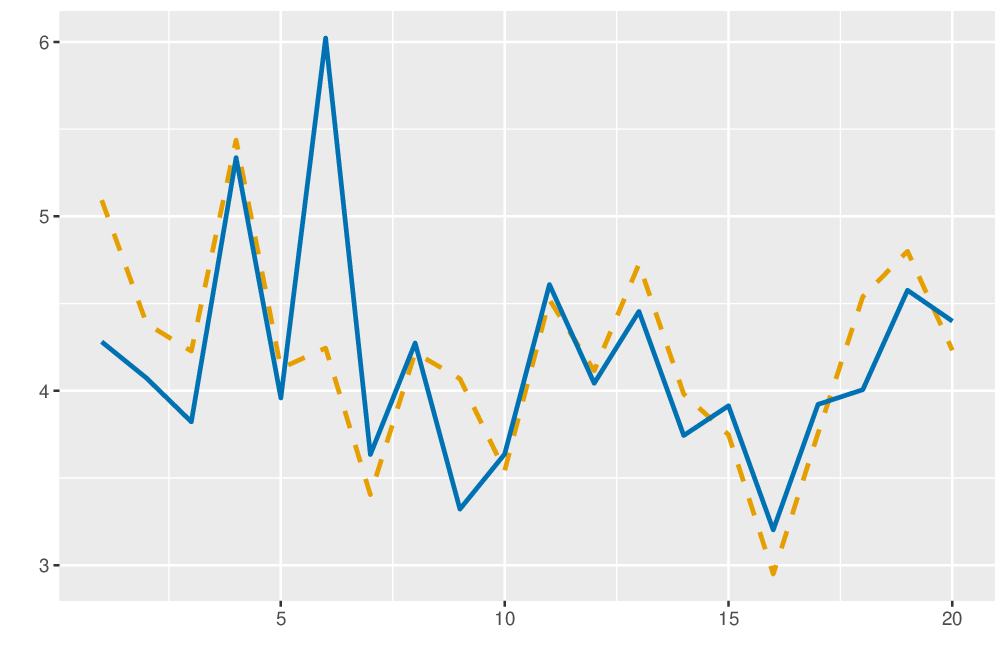}
        \caption{Results for $d$ (yellow) and $d_{\phi}$ (blue).}
    \end{subfigure}
    \begin{subfigure}{0.485\textwidth}
        \centering
        \includegraphics[width=\linewidth]{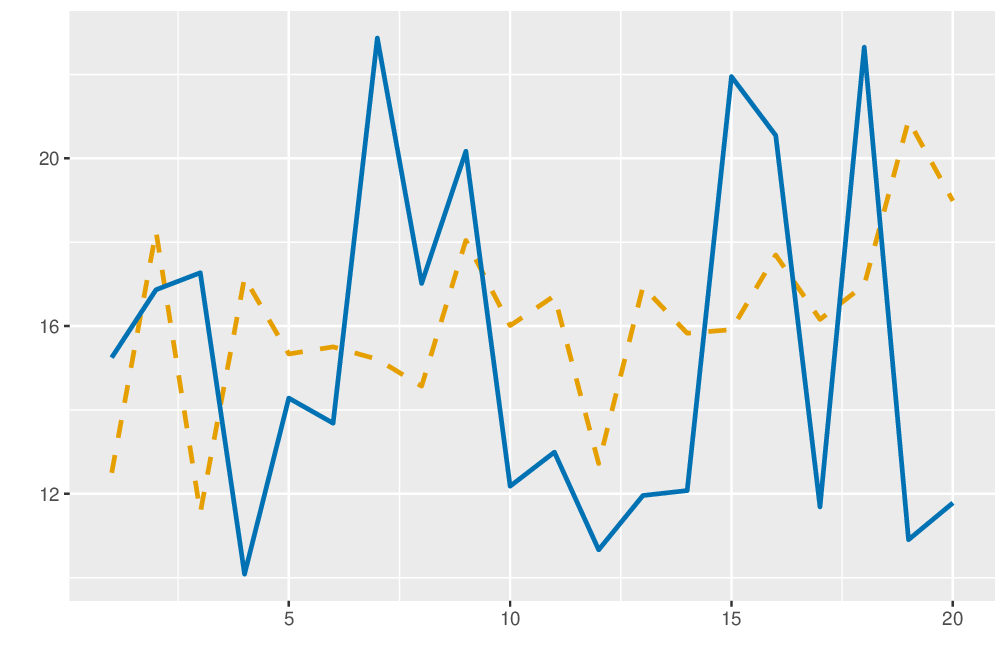}
        \caption{Results for $d$ (yellow) and $d_{\psi}$ (blue).}
    \end{subfigure}
    \caption{Comparison of extension errors using McShane-Whitney formulas, for different metrics.}
    \label{fig:errors}
\end{figure}

We also provide a comparison of these results with those obtained by applying different regression algorithms widely studied in the literature: neural networks and linear regression. The results of this comparison are summarised in Table \ref{tab:nnlr}. It is noteworthy that the results obtained by a neural network are comparable to those obtained by the McShane-Whitney technique, both in terms of the estimated error and its deviation, and in terms of execution time. We can also compare the performance of the standard indices with that of a linear regression. For our example, similar results are obtained, except for the computation time, where linear regression proves to be more efficient. To allow a more comprehensive comparison of these behaviours, Figure \ref{fig:errors} illustrates the different errors obtained in each iteration of the cross-validation. Each of the $\psi$ metric models is plotted next to the most similar one discussed above.

\begin{table}[h!]
\centering
    \begin{tabular}{|c|c|c|c|c|}
    \hline
        &Standard & \tiny McShane-Whitney \normalsize & Neural Net & Linear\\
        \hline
        \hline
        Mean RMSE & 16.69 & 4.55 & 4.40 & 13.80 \\
        \hline
        Median RMSE & 16.50 & 4.47 & 4.42 & 13.65 \\
        \hline
        Standard deviation & 1.52 & 0.63 & 0.46 & 3.25 \\
        \hline
        Seconds per iteration & 3.000$\times 10^{-1}$ & 3.013$\times 10^{-1}$ & 1.923$\times 10^{-1}$ & 4.341$\times 10^{-3}$\\
        \hline
    \end{tabular}
    \caption{Performance of $\psi$-metric models and other regression methods.}
    \label{tab:nnlr}
    \end{table}

\begin{figure}[h!]
    \begin{subfigure}{0.485\textwidth}
        \centering
        \includegraphics[width=\linewidth]{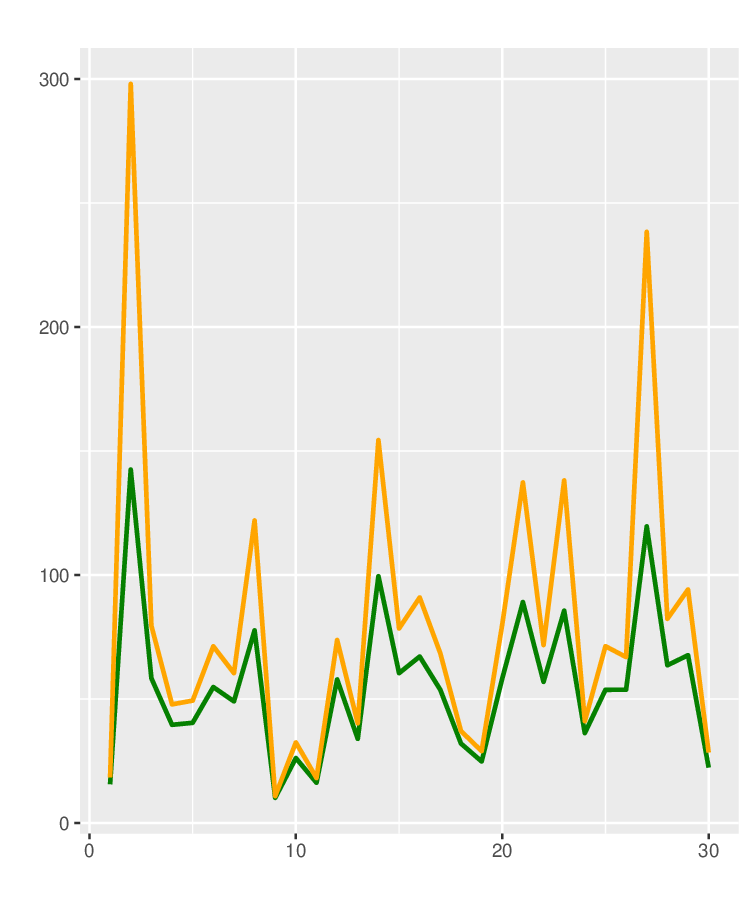}
        \caption{Results for $d$ (yellow) and $d_{\phi}$ (green).}
    \end{subfigure}
    \begin{subfigure}{0.485\textwidth}
        \centering
        \includegraphics[width=\linewidth]{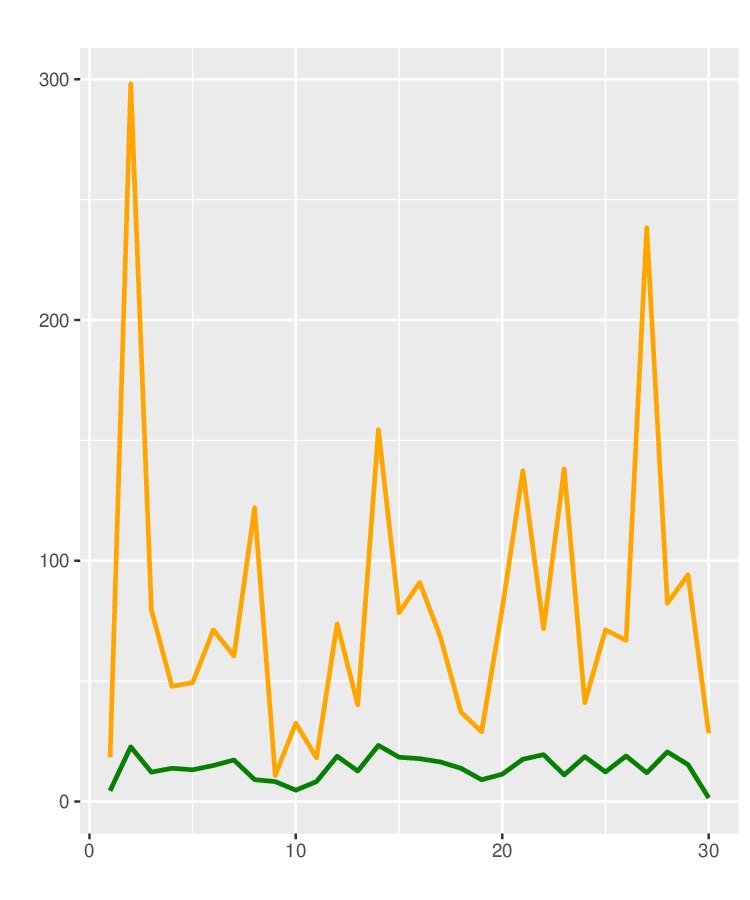}
        \caption{Results for $d$ (yellow) and $d_{\psi}$ (green).}
    \end{subfigure}
    \caption{Comparison of extension errors using standard indices, for different metrics.}
    \label{fig:phipsist}
\end{figure}

\begin{figure}[h!]
    \begin{subfigure}{0.485\textwidth}
        \centering
        \includegraphics[width=\linewidth]{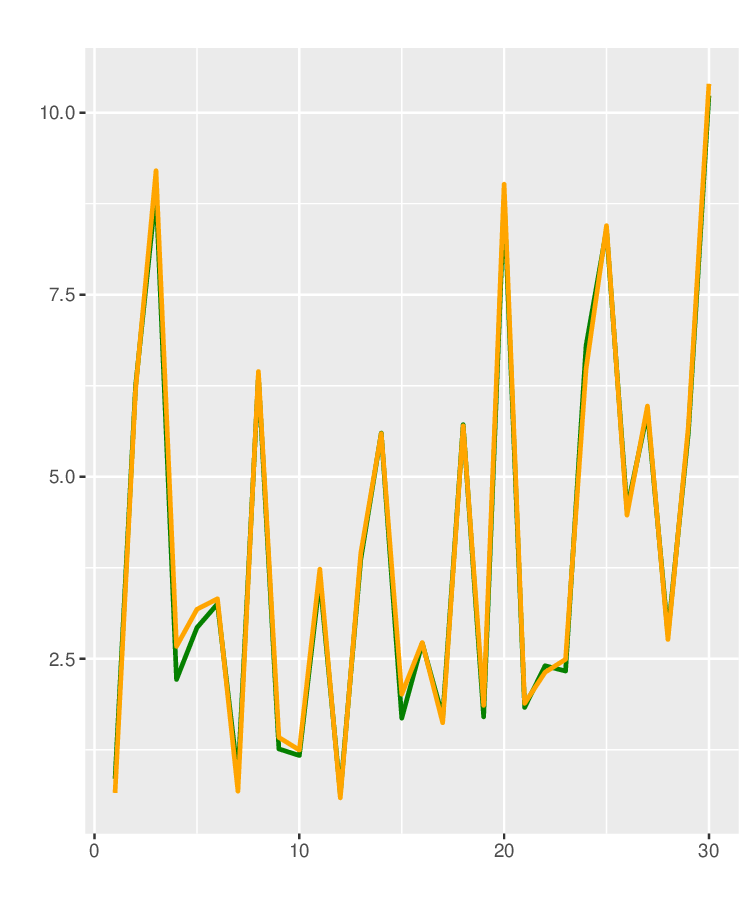}
        \caption{Results for $d$ (yellow) and $d_{\phi}$ (green).}
    \end{subfigure}
    \begin{subfigure}{0.485\textwidth}
        \centering
        \includegraphics[width=\linewidth]{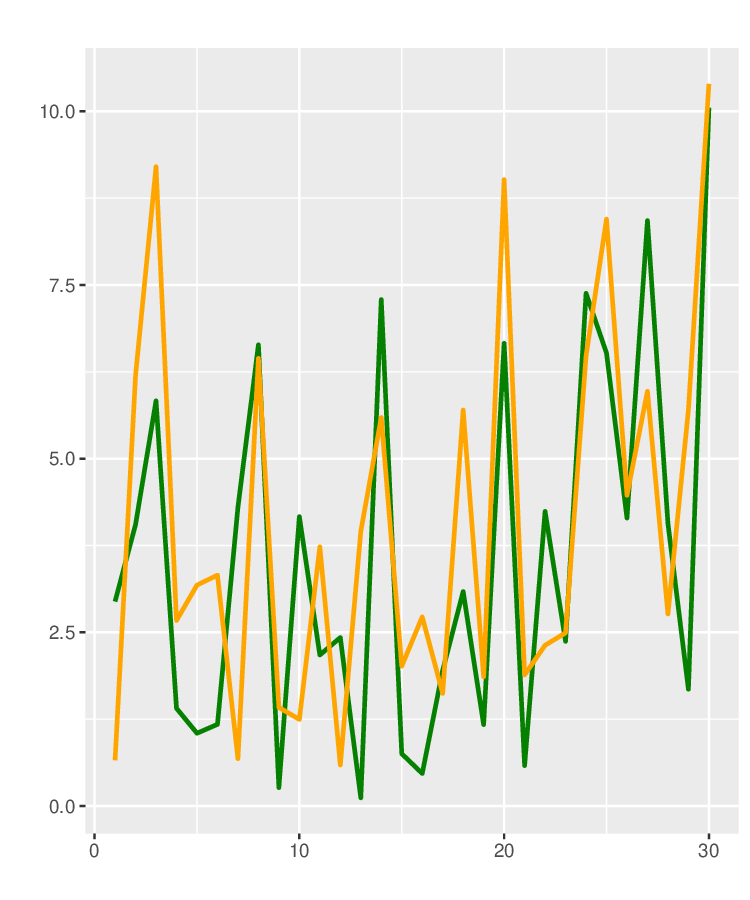}
        \caption{Results for $d$ (yellow) and $d_{\psi}$ (green).}
    \end{subfigure}
    \caption{Comparison of extension errors using McShane-Whitney formulas, for different metrics.}
    \label{fig:phipsimw}
\end{figure}

It should also be noted that the way in which this improvement is achieved could vary from one method to another and from one metric to another. With this in mind, we have taken a concrete division into training and test sets (the 71 most populated cities will serve as training and the rest as test) and applied the results of the extension studied. In the case of standard indices, if we look at Figure \ref{fig:phipsist}, we can see that by redefining the metric to $d_{\phi}$ or $d_{\psi}$, the improvement is general, i.e. we obtain a lower error for each element following the original trend. However, in the case of the McShane-Whitney method, Figure \ref{fig:phipsimw} shows that depending on the new metric, the accuracy can be different for each element, although on average it is lower, as we have seen before.

Finally, in the Table \ref{tab:canrank} we present the predictions resulting from each method and provide a ranking according to them. Since we do not have a benchmark for our index, we will compare it with other existing indices when evaluating the resulting rankings. The Mercer quality of living city ranking\footnote{https://mobilityexchange.mercer.com/Insights/quality-of-living-rankings} classifies Canadian cities in the following order: Vancouver, Toronto, Ottawa, Montreal and Calgary. As we can see, this ranking is reasonably consistent with the one we have offered and, although the standard indices do not provide an exact approximation of the index in question, they do provide an accurate overview when it comes to establishing a ranking. 

   \begin{table}[h!]
    \centering
    \begin{tabular}{|c|c|c|c|c|}
    \hline
        Ranking & Standard & McShane-Whitney & Neural net & Linear \\ \hline \hline
        1 & Vancouver & Montreal & Vancouver & Vancouver \\ 
        2 & Toronto & Vancouver & Toronto & Toronto \\ 
        3 & Montreal & Longueuil & Montreal & Montreal \\ 
        4 & Burnaby & Toronto & Burnaby & Burnaby \\ 
        5 & Longueuil & Saskatoon & Longueuil & Longueuil \\ 
        6 & Mississauga & Winnipeg & Ottawa & Ottawa \\ 
        7 & Winnipeg & Burnaby & Winnipeg & Winnipeg \\ 
        8 & Ottawa & Mississauga & Mississauga & Surrey \\ 
        9 & Brampton & Ottawa & Brampton & Laval \\ 
        10 & Quebec & Brampton & Quebec & Mississauga \\ 
        11 & Surrey & Surrey & Laval & Kitchener \\ 
        12 & Laval & Quebec & Surrey & Brampton \\ 
        13 & Kitchener & Edmonton & Kitchener & Hamilton \\ 
        14 & Calgary & Kitchener & Calgary & Saskatoon \\ 
        15 & Saskatoon & Windsor & Gatineau & Calgary \\ 
        16 & Markham & Laval & Markham & Quebec \\ 
        17 & Hamilton & Hamilton & London & Windsor \\ 
        18 & Edmonton & Calgary & Hamilton & Edmonton \\ 
        19 & London & London & Edmonton & Vaughan \\ 
        20 & Gatineau & Gatineau & Windsor & Markham \\ 
        21 & Vaughan & Markham & Vaughan & London \\ 
        22 & Windsor & Vaughan & Saskatoon & Gatineau \\ \hline
    \end{tabular}
    \caption{Ranking of Canadian cities}
    \label{tab:canrank}
\end{table}

\section{Conclusions}
In this paper we have shown a new methodology for the extension of indices defined on metric subspaces of metric models. The main contribution of our results is that they include a new class of metrics (composition metrics) that give us more flexibility in the approximation tools for finding adapted results of Lipschitz regression processes. While the state of the art Lipschitz extensions are highly dependent on the chosen metric, this methodology provides us with a straightforward way to improve extension results without a complex study of which metrics are well suited to the problem. In particular, the introduction of this wide class of metrics starting from a standard distance (e.g. the Euclidean distance) allows this improvement by reducing the Lipschitz constant. We provide the approximation formulas as well as the error bounds for the proposed procedure.

As an applied example, we show how we can extend the so-called liveability index for large urban centres in the United States, for which this index is known, from Canadian cities. These results are interesting in themselves as they provide more information on a topic of current interest. The results show that our techniques produce similar results to other widely used techniques, with the advantage of better interpretability.

\vspace{6pt}

\paragraph{\bf Contribution}
Conceptualization, R.A., A.G., and E.A.S.P.; formal analysis, R.A. and E.A.S.P.; investigation, J.M.C., A.G. and E.A.S.P.; methodology, R.A. and A.G.; supervision, J.M.C.; writing original draft, A.G. and E.A.S.P.; writing—review and editing, R.A. and J.M.C. All authors have read and agreed to the present version of the manuscript.

\paragraph{\bf Funding}
Enrique A. Sánchez Pérez was supported by
Grant PID2020-112759GB-I00 funded by MCIN/AEI /10.13039/501100011033. R. Arnau was supported by a contract of the Programa de Ayudas de Investigación y Desarrollo (PAID-01-21), Universitat Politècnica de València.

\paragraph{\bf Data availability}
All the algorithms and visualizations of this paper can be found in the link of GitHub \url{https://github.com/Algoncor/Composition-metrics}.

\bibliographystyle{plain}
\bibliography{biblio}

\end{document}